\documentclass[%
 aip,
 amsmath,amssymb,
 reprint,%
]{revtex4-2}

\usepackage{graphicx}
\usepackage{dcolumn}
\usepackage{bm}

\usepackage[utf8]{inputenc}
\usepackage[T1]{fontenc}
\usepackage{mathptmx}
\usepackage{etoolbox}
\usepackage{float}
\usepackage{amsmath}
\usepackage{amssymb}
\usepackage{xcolor}

\newcommand{\be}{\begin{equation}}
\newcommand{\ee}{\end{equation}}
\newcommand{\bea}{\begin{eqnarray}}
\newcommand{\eea}{\end{eqnarray}}

\newcommand{\la}{\langle}
\newcommand{\ra}{\rangle}

\newcommand{\dd}{\delta}
\newcommand{\eps}{\varepsilon}
\newcommand{\dm}{\mathrm{d}}

\makeatletter
\def\@email#1#2{%
 \endgroup
 \patchcmd{\titleblock@produce}
  {\frontmatter@RRAPformat}
  {\frontmatter@RRAPformat{\produce@RRAP{*#1\href{mailto:#2}{#2}}}\frontmatter@RRAPformat}
  {}{}
}%
\makeatother
\begin{document}

\preprint{AIP/123-QED}

\title{Mixed-Mode Chimera States in Pendula Networks}
\author{P. Ebrahimzadeh}
 \affiliation{Forschungszentrum J\"{u}lich GmbH, ZEA-2: Electronics Systems, 52428 J\"{u}lich, Germany}
 \email{p.ebrahimzadeh@fz-juelich.de.}
\author{M. Schiek}%
 \affiliation{Forschungszentrum J\"{u}lich GmbH, ZEA-2: Electronics Systems, 52428 J\"{u}lich, Germany}
 \email{m.schiek@fz-juelich.de.}
\author{Y. Maistrenko}
 \affiliation{Forschungszentrum J\"{u}lich GmbH, ZEA-2: Electronics Systems, 52428 J\"{u}lich, Germany}
 \affiliation{Institute of Mathematics and Technical Centre, NAS of Ukraine, Tereschenkivska Str. 3, 01030, Kyiv, Ukraine }
 \email{y.maistrenko@biomed.kiev.ua}

\date{\today}

\begin{abstract}
We report the emergence of peculiar chimera states in networks of identical pendula with global phase-lagged coupling. The states reported include both rotating and quiescent modes, i.e. with non-zero and zero average frequencies. This kind {\it mixed-mode chimeras} may be interpreted as images of bump states known in neuroscience in the context of modelling the working memory. We illustrate this striking phenomenon for a network of $N=100$ coupled pendula, followed by a detailed description of the minimal non-trivial case of $N=3$.  Parameter regions for five characteristic types of the system behavior are identified consisting: two mixed-mode chimeras with one and two rotating pendula, classical weak chimera with all three pendula rotating, synchronous rotation and quiescent state. The network dynamics is multistable: up to four of the states can coexist in the system phase state as demonstrated through the basins of attraction. The analysis suggests that the robust mixed-mode chimera states can generically describe the complex dynamics of diverse pendula-like systems widespread in nature.
\end{abstract}

\maketitle

\begin{quotation}
Chimera states generally refer to spatio-temporal patterns in networks of identical or close to identical oscillators, in which a group of oscillators is synchronized and the other group is asynchronous. For networks composed of Kuramoto oscillators with inertia, chimera states are manifested in the form of {\it solitary states} in which one or a few oscillators split off from the main synchronized cluster and start to rotate with a different average frequency. Chimeras of this kind include rotational modes and their frequencies are determined by the system parameters. In networks of excitable elements such as neurons, in contrary, chimeric spatiotemporal patterns typically arise in the form of {\it bump states}, where active spiking neurons (large amplitude) coexist with quiescent (subthreshold) ones. The bumps states are created due to the competition mechanism between attractive and repulsive couplings, which suppresses the quiescent group. Then, the pendulum network can be viewed as a model bringing together the properties of the Kuramoto oscillators with inertia and the excitable theta neuron model, for which we show the emergence of mixed-mode chimeras with non-zero and zero average frequencies of individual oscillators from different groups.
\end{quotation}

\section{\label{intro} Introduction. Mixed-Mode Chimera States}

Patterns of synchronization have been the subject of intensive study in various fields, ranging from biology, social behavior and network science among others \cite{kuramoto2003chemical, pikovsky_rosenblum_kurths_2001, strogatz2001exploring}. To this end, "classical" chimera states refer to spatio-temporal patterns emerging in networks of non-locally coupled oscillators as coexistence of coherent and incoherent groups \cite{Panaggio_2015, scholl2016synchronization, omel2019chimerapedia, zakharova2020chimera, parastesh2021chimeras}.
In recent works, the notion of chimera states has been generalized to the property of frequency clustering, named weak chimera states \cite{ashwin2015weak}. Among weak chimeras a distinctive role is played by so-called {\it solitary states} \cite{jaros2018solitary, berner2020solitary, jaros2021chimera, schulen2021solitary, hellmann2020network, franovic2022unbalanced, maistrenko2014solitary, maistrenko2020spiral, maistrenko2020chimeras} in which one or a few (or even more) oscillators split off from the main synchronized cluster and start rotating with a different average frequency. Characteristic examples of this kind behavior are supplied by the Kuramoto model with inertia \cite{jaros2015chimera, ermentrout1991adaptive, olmi2015intermittent, olmi2015chimera, belykh2016bistability, belykh2017foot, brister2020three, kruk2020solitary}, where solitary states arise at all types of the network coupling from global to local \cite{jaros2018solitary}. An essential property of the solitary states is that the desynchronized oscillators do not create localized groups in the space, in contrast to the classical Kuramoto model without inertia \cite{kuramoto2002coexistence, PhysRevLett.93.174102}. Instead, the splitted oscillators appear to be distributed in the network space in visually arbitrary manner subjected to the assigned initial conditions. This fact causes a huge multiplicity of the coexisting stable solitary states of different configurations, given not only by different number of splitted elements but also their permutation in the space. The number of coexisting solitary states grows exponentially and network shows spatial chaos in
the thermodynamic limit \cite{omelchenko2011loss}.
\begin{figure*}
            \includegraphics[width=\linewidth]{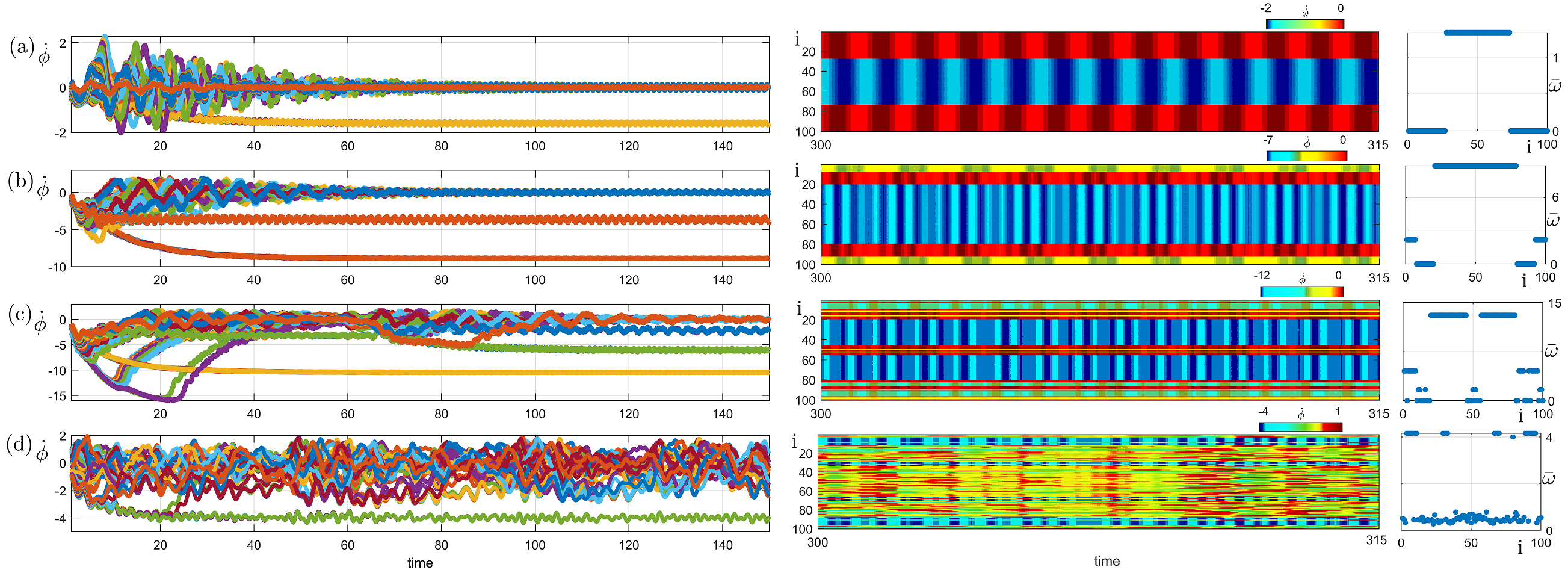}
	  \caption{ \label{fig11} Typical examples of mixed-mode chimera state with different number of frequency clusters for model~(\ref{PN}) of $N=100$ pendula. Left panel:  velocity time plots for $\mu = 2$ and $\alpha = 1.2$(a), $1.35$(b), $1.42$(c),$1.52$(d), respectively, and with the same fixed initial conditions. Middle panel: sustained frequency profiles of the mixed-mide chimeras (inter-cluster dynamics). Right panel: average frequencies for the individual pendula of the mixed-mode chimeras. Other parameters: $m=1.0, \eps=0.1, r = 1, \omega = 0$.}
\end{figure*}

In this work, we consider a network of globally coupled pendula
\be
\label{PN}
m\ddot\phi_i + \eps \dot \phi_i + r \sin\phi_i= w + \frac{\mu}{N} \sum_{j=1}^{N} \sin[\phi_j - \phi_i - \alpha],
\ee
where $\phi_i$, $i=1,...,N$, are phase variables, $\mu$ is the coupling strength, $\alpha$ is the phase-lag, $m$, $\eps$ and $w$ are inertia, damping, and natural frequency of each pendulum, respectively. Experimental evidence of chimera states in the pendula network~(\ref{PN}) was obtained for a ring setup of coupled metronomes \cite{kapitaniak2014imperfect, wojewoda2016smallest, ebrahimzadeh2020minimal}. We find that, besides the amazing chimera complexity \cite{brezetsky2021chimera} inherent to the Kuramoto model with inertia, model~(\ref{PN}) also gains new characteristic solutions caused by the presence of the nonlinear gravitation terms $\sin\phi_i$. We name them {\it mixed-mode} chimera states, in which a part of oscillators are in a quiescent mode (slightly oscillating, however) and the others are rotating with a non-zero average frequency. Note that, behavior of this type is widespread in neuroscience, known as {\it bump states}\cite{Ermentrout_1998, laing2001stationary, laing2002multiple, owen2007bumps, laing2021interpolating} which combines the spatially localized groups of persistent neuronal activity at the silent background of not-firing subthreshold neurons. In particular, bump states are considered as appropriate models for functioning of the working memory in the brain \cite{RENART2003473, coombes2005waves, MILLER2018463}. During the last two decades, bump states were intensively studied for various excitable neuronal models, including their apparent relation to chimera states \cite{Ermentrout_1998, laing2001stationary, laing2002multiple, owen2007bumps, laing2021interpolating}. 

The emergence of the mixed-mode chimera states in model~(\ref{PN}) is manifested as the mismatch in average frequency between the chimera clusters. One of the clusters is similar to the quiescent background in bumps, i.e., with zero average frequency, and the other one consist of rotating oscillators with non-zero average frequency. An important characteristic is that to induce the mixed-mode chimera of a given configuration, specially prepared initial conditions are needed (similarly again to the bump states). In the dynamical interpretation, this property follows from the complex structure of the basins of attraction, in particular, due to fractal basin boundary.

The dynamics of a single pendulum in model~(\ref{PN}) consists of a stable equilibrium and stable limit cycle including their coexistence at some parameter region \cite{belykh2016bistability}. For small values of $w$ (and $\eps$), the fixed point solution exists and is stable. Increasing $w$ results in the appearance of a limit cycle which is born in a homoclinic bifurcation at the Tricomi bifurcation curve \cite{tricomi1933integrazione}. With further increase of $w$ at line $w/r = 1$, the stable fixed point is eliminated in a saddle-node bifurcation, and the only attractor is the limit cycle. In the bistability region between the Tricomi curve and $w/r=1$, both fixed point and limit cycle can develope depending on the initial conditions. We note that, at $r=0$ the model transforms into the Kuramoto model with inertia, the stable equilibrium is eliminated and dynamical regimes reduce to that of limit cycle with average frequency $\la \dot\phi \ra = w / \eps$. In our numerical simulation we fix parameters $m=1$, $\eps = 0.1$, $r=1$, and setting the natural frequency $w = 0$. The complex chimera-like regimes in the model~(\ref{PN}) arise due to the influence of coupling, which include self-coupling playing the role of an "external forcing".
\section{\label{n100} Network $N = 100$}

Typical examples of mixed-mode chimeras in model~(\ref{PN}) of $N=100$ pendula with global coupling are illustrated in Fig~\ref{fig11}. The figure reveals the appearance of solutions with different number and size of the frequency clusters, where one of the clusters is quiescent (only slightly oscillating) but the others are rotationg with non-zero average frequency. The solutions are obtained by the simulations starting from the same initial conditions for increasing  values of the phase-lag parameter $\alpha=1.2(a); 1.35(b); 1.42(c); 1.52(d)$, where the coupling strength is fixed to $\mu=2$. The left panel in the figure shows the transient behavior of the solutions, middle panel emboldens the shape of the sustained cluster oscillations (after the transient) within a narrow time interval, and the average frequencies are depicted in the right panel. It can be seen that the complexity of the mixed-mode chimeras grows with the increase of $\alpha$. First, in Fig.~\ref{fig11}(a), the frequency profile is rather simple, resembling classical bump state: some pendula of the network rotate, the rest are quiescent. The number of the rotating clusters grows with an increase of $\alpha$: two such clusters can be seen in (b) and three in (c). With further increase of $\alpha$ the dynamics inside the quiescent cluster becomes irregular or even chaotic, showing a "fuzzy" structure of the individual frequencies, see (d).

Each cluster in a mixed-mode chimera state is categorized by a common averaged frequency of the oscillators $\bar \omega$, with $\bar \omega \ne 0$ for rotating and $\bar \omega = 0$ for oscillating cluster in the background. This is in contrast to classical chimera states, where a synchronous rotating cluster coexists with a group of incoherent oscillators with a bell shape frequency profile. Furthermore, the mixed-mode chimeras are not chaotic transients as the classical counterpart \cite{wolfrum2011chimera}, i.e. do not collapse into a synchronization or rotating wave states. In contrary, they are persistent solutions similar to solitary states in the Kuramoto model with inertia \cite{jaros2018solitary}.
 
Note that, frequency clustering in chimera states does not imply, in general, phase clustering  (see Ref.~[\onlinecite{maistrenko2017smallest}] for illustrative examples). In the model~(\ref{PN}), however, this is the case as soon as  $\alpha<\pi/2$: after formation of a mixed-mode chimera, not only frequencies but also phases coincide for the oscillators within each cluster. The dynamics is then governed on an invariant manifold of the reduced dimension 
\bea
\label{DM}
D(M) = \{ \phi_1 = ... = \phi_{n_1} = \Phi_1(t);    ...;  \phi_{n_M} =        ... = \phi_N = \Phi_M(t); \nonumber\\
\dot\phi_1 = ... = \dot\phi_{n_1} = \dot\Phi_1(t); ...;  \dot\phi_{n_M} = ... = \dot\phi_N = \dot\Phi_M(t) \} \nonumber\\
\eea 
such that the in-manifold dynamics is given by the equations
\be
\label{PM}
m\ddot\Phi_i + \eps \dot \Phi_i + r \sin(\Phi_i)= -n_i \tilde{w} + \mu \sum_{j \ne i}^{M} n_j \sin[\Phi_j - \Phi_i - \alpha],
\ee
where $\tilde w = \mu \sin(\alpha)$ is the modified eigenfrequency and $n_i$ is the ratio of pendula in the $i$'s cluster to the network size. For different values of parameters $(\alpha, \mu)$ and the initial conditions, Eq.~(\ref{PM}) can develope in different dynamical regions (such as fixed point, bistability, limit cycle, invariant torus or chaos).  

An analytical approach with asymptotic description can be developed to describe the cluster dynamics. One can represent $\Phi_i$'s as sum of a monotonic term with respect to time and correction terms scaled by small perturbation parameter $\lambda$ 
\be
\label{pert}
\Phi_i = \bar\omega_i t + \sum_{l=1}^{\infty} \lambda^{l} \delta \Phi^{l}_i
\ee
Inserting condition~(\ref{pert}) into Eq.~(\ref{PM}) and sorting by $\lambda$, one finds (Appendix~\ref{AP1}) the average frequency of the rotating clusters 
\be
\label{wm}
\bar \omega_i = -\frac{n_i \tilde{w}}{\eps}
\ee
and the leading correction term
\be
\label{CR1}
\lambda\delta \Phi^{1}_i \sim  \frac{r}{m \bar\omega_i ^2} \sin\bar\omega_i t - \mu \sum_{j=1}^{M} \frac{n_j}{\Delta \bar\omega_{ji} ^2} \sin[\Delta \bar\omega_{ji} t - \alpha],
\ee
where $\Delta \bar\omega_{ji} = \bar\omega_j - \bar\omega_i$. The average frequency of the clusters depend on the parameters $(\alpha, \mu)$ and the number of pendulum in the cluster. The system dynamic from this point of view for any $N$ could be a subject of future study. In the next section, we consider the emergence and coexistence of mixed-mode chimeras for the minimal but non-trivial network of $N=3$ coupled pendulua in model~(\ref{PN}).
\section{\label{N3}Minimal Network $N = 3$}

The minimal network with chimeric behavior in model~(\ref{PN}) consists of $N=3$ coupled pendula
\bea
\label{n31}
\ddot \phi_1 + \eps \dot \phi_1 + \sin{\phi_1} &=& \frac{\mu}{3} [ \sin(\phi_2 - \phi_1-\alpha)+ \sin(\phi_3 - \phi_1- \alpha) - \sin{\alpha} ], \nonumber\\
\ddot \phi_2 + \eps \dot \phi_2 + \sin{\phi_2} &=& \frac{\mu}{3} [ \sin(\phi_1 - \phi_2 - \alpha)+\sin(\phi_3 - \phi_2-\alpha) - \sin{\alpha} ], \nonumber\\
\ddot \phi_3 + \eps \dot \phi_3 + \sin{\phi_3} &=& \frac{\mu}{3} [ \sin(\phi_1 - \phi_3 - \alpha)+\sin(\phi_2 - \phi_3-\alpha) - \sin{\alpha} ], \nonumber\\
\eea
where we put $m=r=1$ and $w=0$ for simplicity. In this minimal system, we find two types of mixed-mode chimera states with one and two rotating pendula, respectively. We symbolically denote them by the average frequencies $(\bar\omega_1, \bar\omega_2, \bar\omega_3)$ of the pendula such that $(0:1:1)$ corresponds to the mixed-mode chimera with two rotating pendula and one oscillating, and $(0:0:1)$ denotes one rotating pendulum and two oscillating. By $(1:2:2)$ we denote a "classical" chimera in which all pendula are in a rotation mode such that two are synchronized and rotating faster than the third one. In addition, there are two non-chimeric behaviors: synchronous rotating state $(1:1:1)$ and trivial equilibrium $(0:0:0)$ given by in-phase fixed point $\phi_1=\phi_2=\phi_3$.

\begin{figure}
\includegraphics[width=\columnwidth]{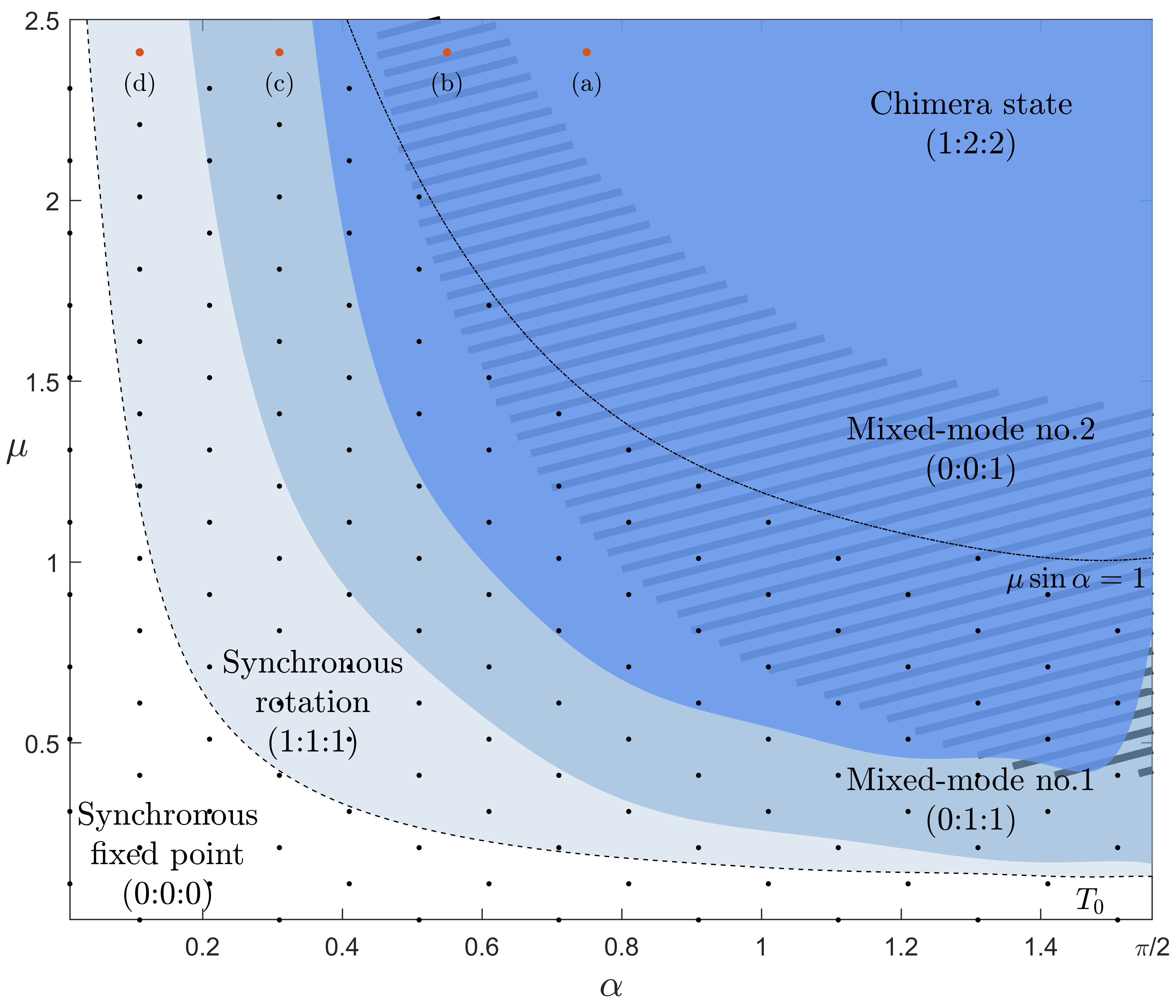}
\caption{\label{fig2} Intersecting regions of chimera states for system~(\ref{n31}) of $N=3$ coupled pendula in the ($\alpha, \mu$) parameter plane.  The chimeras reside on wide areas of the parameter plane. Type-1 mixed-mode chimera (0:1:1) (light blue region) extends for large values of $\mu$ for all $\alpha$, while the type-2 mixed-mode chimera (0:0:1) only exists in the strip region (shown dashed). Region for classical chimera (1:1:2)  is shown in dark blue, it is also extends for large $\mu$. Synchronous rotation state (light gray) emerges in a homoclinic bifurcation at the Tricomi curve $T_0$; it exist and is stable for all values of $(\alpha, \mu)$ above $T_0$. quiescent fix point state (trivial equilibtium) exists for small $\alpha$ and $\mu$ until the saddle-node bifurcation curve $\mu \sin{\alpha}=1$.  Characteristic example of the patterns are presented in Fig.~\ref{fig3}}
\end{figure}

Results of direct numerical simulation of the system~(\ref{n31}) in the two-parameter plane of the phase-lag $\alpha$ and coupling strength $\mu$ are presented in Fig.~\ref{fig2}. This figure reveals the appearance of regions of the chimera states of the three types $(0:1:1)$, $(0:0:1)$, and $(1:2:2)$, all arising at not-small values of $\alpha$ and $\mu$. Alternatively, if $\alpha$ and/or $\mu$ are close to zero, the network displays only the equilibrium state $(0:0:0)$, which exists and is stable in the dotted region below the saddle-node bifurcation curve $\mu\sin\alpha=1$.  
\begin{figure}[b!]
\includegraphics[width=\columnwidth]{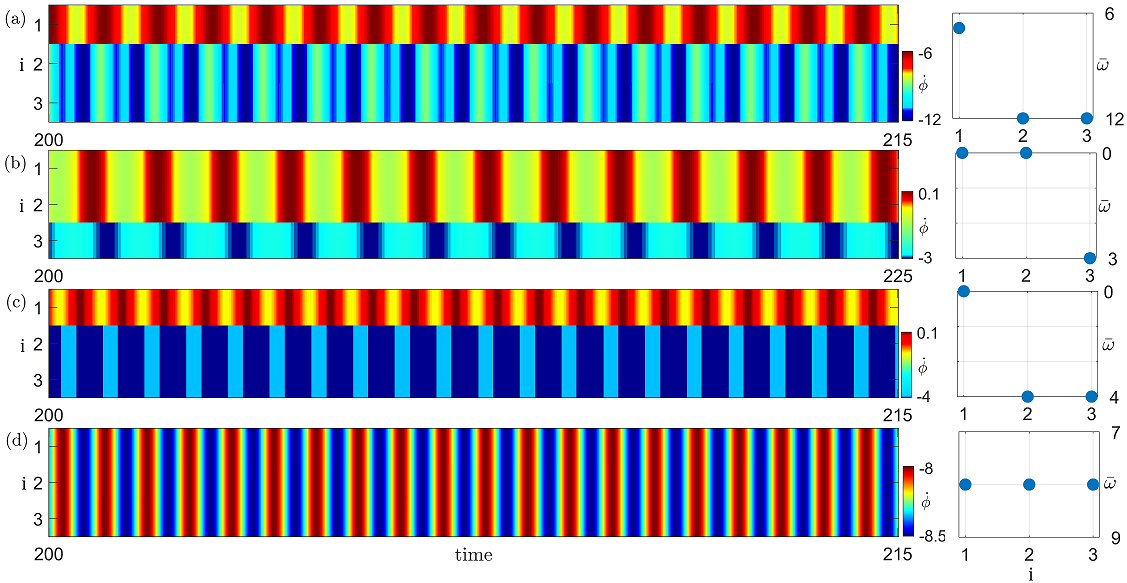}
\caption{\label{fig3} Frequency time plots of typical behaviors in system~(\ref{n31}) of $N=3$ coupled pendula. Setting coupling strength $\mu=2.4$, the following patterns are observed for different values of $\alpha$: (a) $\alpha = 0.8$: classical chimera state $(1:2:2)$ ; (b) $\alpha = 0.6$: type-1 mixed-mode chimera $(0:0:1)$ with two quiescent and one rotating pendula; (c) $\alpha = 0.4$: type-2 mixed-mode chimera $(0:1:1)$ with one quiescent and two rotating pendula; and (d) $\alpha = 0.2$: synchronous rotation $(1:1:1)$. Other parameters as in Fig.~\ref{fig2}.}
\end{figure}
\begin{figure}
\includegraphics[width=\columnwidth]{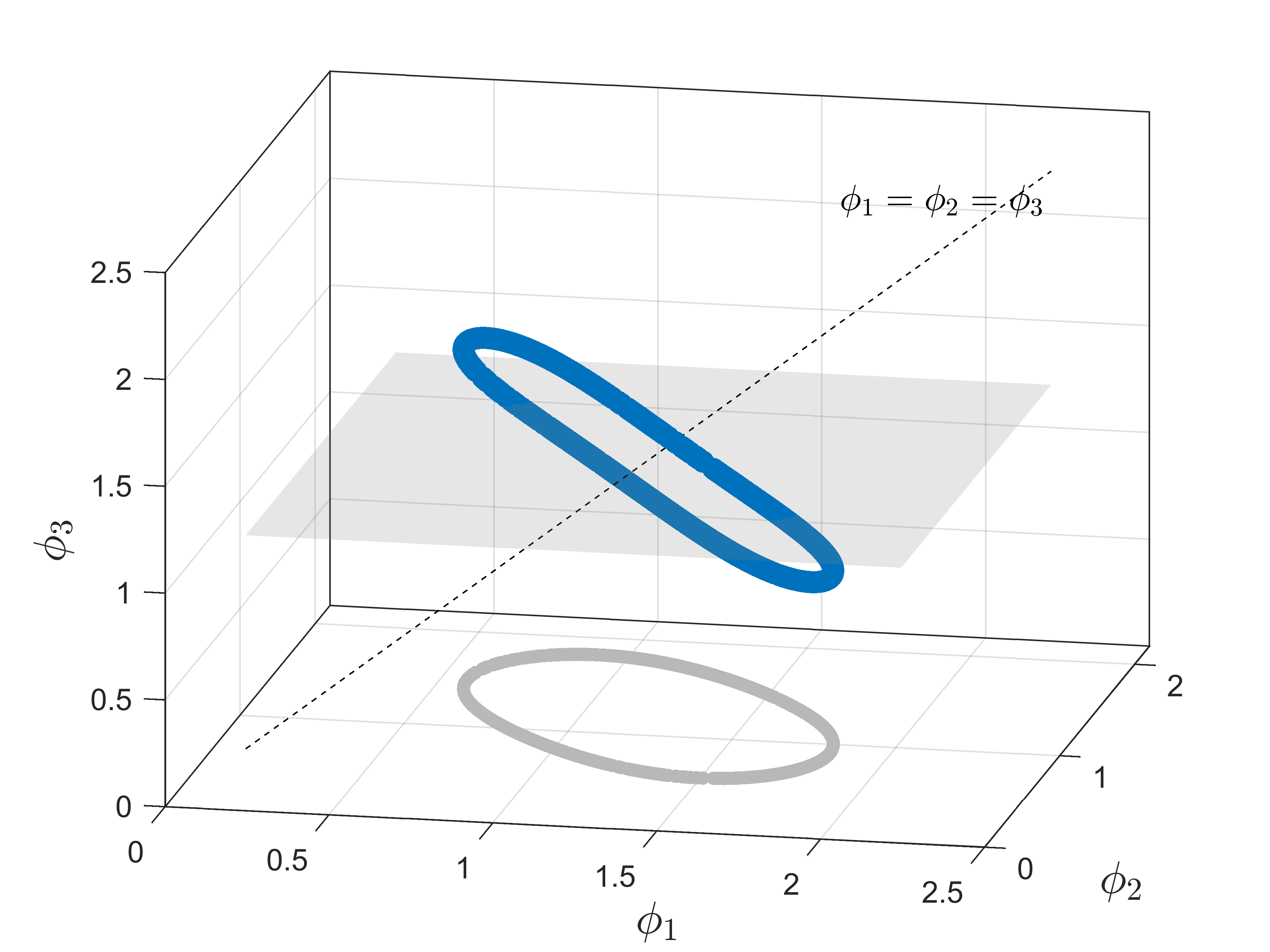}
\caption{\label{fig5} Ring of unstable asynchronous fixed points of system~(\ref{n31}) of $N=3$ coupled pendula at parameters $\alpha = 1.2$ and $\mu = 1.2$}
\end{figure}

Note that both, the mixed-mode chimera $(0:1:1)$ and the rotating chimera $(1:2:2)$ reside in large parameter regions with the area of coexistence (shown in blue and dark gray, respectively), and they extend up to the large values of $\alpha$ and $\mu$. The second mixed-mode chimera $(0:0:1)$, in contrary, exists in a smaller region (shown in dashed). Non-chimeric synchronous rotating state $(1:1:1)$ also occupies a big parameter region. It arises in a homoclinic bifurcation at the curve $T_0$ (equivalent to the Tricomi curve) and is stable for all $\alpha, \mu$ above $T_0$.

To describe the mechanism of the bifurcations in system~(\ref{n31}) of $N=3$ pendula, we start by identifying its fixed points in the attractive parameter region $\alpha < \pi/2$. Note that system~(\ref{n31}) represents a six-dimensional system of differential equations, and due to the presence of the nonlinear gravitation term $\sin \phi$, cannot be reduced to a 4-Dim system in phase differences (contrary to the corresponding Kuramoto model with inertia) 

System~(\ref{n31}) has a $2D$ invariant synchronous manifold $\mathfrak{M} = \{\phi_1=\phi_2=\phi_3 \}$, its dynamics is given by a single pendulum equation 
\be
\label{p1}
\ddot\phi_i + \eps \dot \phi_i + \sin(\phi_i)= -\mu \sin\alpha.
\ee
Note that the forcing term in the right hand side of Eq.~(\ref{p1}) arises due to the phase-lagged character of the global coupling, including the self-coupling. System~(\ref{n31}) has two fixed points (equilibria) $O$ and $S$ inside the synchronous manifold $\mathfrak{M}$. In the variables $\{\phi;\dot\phi\}$ they are written as: $O =\{\phi^{*}; 0)\}$ and  $S=\{\phi^{**}; 0)\}$, where
\bea
\label{o12}
\phi^{*} &=&  -\sin^{-1} [\mu \sin\alpha] \\
\phi^{**} &=& \pi +  \sin^{-1} [\mu \sin\alpha].
\eea
\begin{figure*}
            \includegraphics[width=\linewidth]{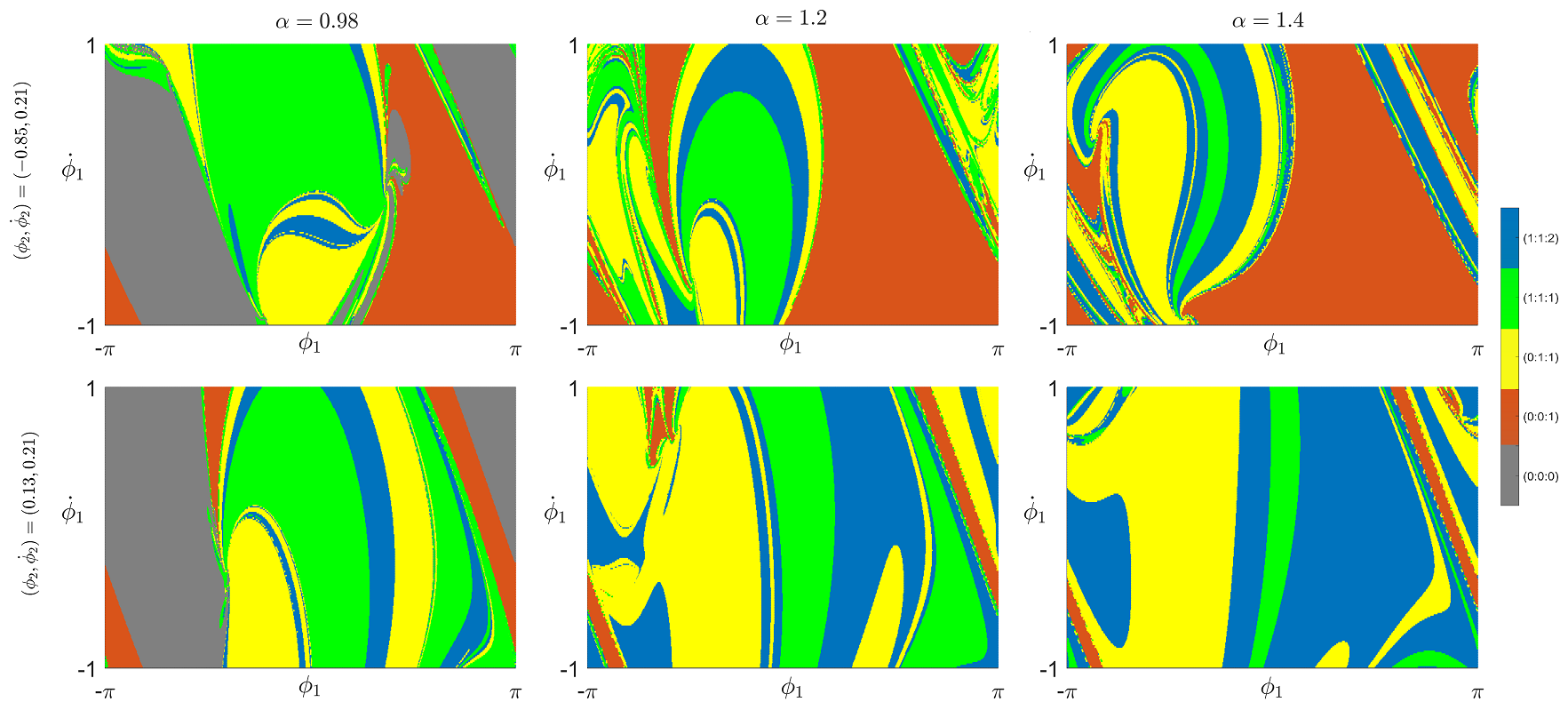}
	 \caption{ \label{fig4} Basins of attraction of system~(\ref{n31}) with $N=3$ for $\mu=1.2$ and increasing  values of $\alpha=0.98; 1.2$; and $1.4$. Top and bottom panels are different cross sections of the six dimensional phase space, fixing different initial conditions for pendulum no.2 (shown in the figure), and the same initial values for pendulum no.3 as $(\phi_3, \dot\phi_3) = (-0.85, 0.21).$ Color gamma of five coexisting characteristic states are shown to the right.}
\end{figure*}
Fixed points $O$ and $S$ exist in the ($\alpha,\mu$) parameter region below the bifurcation curve $\mu=1/\sin\alpha$, at which they collide and disappear in a saddle-node bifurcation. The in-manifold stability of the fixed points is controlled by the characteristic equation for two Lyapunov exponents $\lambda=\lambda_{1,2}$:
\be
\lambda^2 +\eps\lambda \pm \sqrt{1 - \mu^2 \sin^{2}}=0,
\ee
where $(+)$ and $(-)$ stay for $O$ ans $S$, respectively. Then, $S$ is a saddle and $O$ is a stable fixed point such that
\begin{equation}
O:
    \begin{cases}
      \text{stable focus}, &  0<\mu < \sqrt{1 - \frac{\eps^4}{16}} \frac{1}{\sin\alpha} \\
      \text{stable node}, &   \sqrt{1 - \frac{\eps^4}{16}} \frac{1}{\sin\alpha} < \mu < \frac{1}{\sin\alpha}.
    \end{cases}
\end{equation}
Simple analytics confirm that fixed point $O$ is stable not only inside the synchronous manifold $\mathfrak{M}$, but also in transverse directions. Transverse stability of $O$ is obtained from the equation for transverse Lyapunov exponents $\lambda=\lambda_{3-6}$:
\be
(\lambda^2 +\eps\lambda + \sqrt{1 - \mu^2 \sin^{2}}+ \mu\cos\alpha)^{2}=0.
\ee
It has two pairs of two-multiple roots, all having negative real parts (as $\mu\cos\alpha>0$ at $\alpha<\pi/2$), which guaranties transverse stability of $O$. Similar analytics allows to extend this property to larger number of coupled pendula, revealing that $O$ is a stable equilibrium of model~(\ref{PN}) at any finite $N$. The stability follows from the characteristic equation for the eigenvalues $\lambda=\lambda_i, i=1,...,2N$ of $O$:
\be
\label{c1}
(z + \sqrt{1 - \mu^2 \sin^{2} {\alpha}}) (z + \sqrt{1 - \mu^2 \sin^{2} {\alpha}} + \mu \cos{\alpha})^{N-1} = 0,
\ee
where $z=\lambda(\lambda+\eps)$. The first term in Eq.~(\ref{c1}) controls the in-manifold stability and the second one provides two $(N-1)$-multiple transverse eigenvalues of $O$, all characterized by negative real parts at $\alpha<\pi/2$. 

Besides the two synchronous equilibria $O$ and $S$, system~(\ref{n31}) has numerous asynchronous fixed points in the region $\mu > 1$ which, however, are unstable at $\alpha<\pi/2$. A remarkable example is illustrated in Fig.~\ref{fig5}: a continuum of asynchronous fixed points are assembled in a one-dimensional ''ring'' type manifold. All points are unstable for $\alpha<\pi/2$, however, they can stabilize in the repulsive region $\alpha>\pi/2$. Interestingly, a similar situation was reported recently for the $N=4$ dimensional Kuramoto model with inertia, where a continuum of so-called antipodal fixed points exists \cite{brezetsky2021chimera}, unstable at $\alpha<\pi/2$ but stabilizing at $\alpha>\pi/2$.  We leave this issue for future study.  

A typical scenario of the multistable basin transition in system~(\ref{n31}) is illustrated in Fig.~\ref{fig4} for growing values of the parameter $\alpha$, with fixed coupling strength $\mu=1.2$. Two different cross sections of the 6-dimensional phase space are shown in upper and lower panels, respectively, by fixing the initial conditions of pendula no.$2~\&~3$.  The plots demonstrate a highly developed multistability of the system~(\ref{n31}). At $\alpha=0.98$ (left panel) all five characteristic states coexist and occupy wide areas of the space. Here, the parameter point $(\alpha, \mu)$ lies below the curve $\mu \sin{\alpha} =1$. Beyond the curve (middle and right panels), synchronous equilibrium $(0:0:0)$ disappears, and basins of only four states are presented. Noticeably, at different cross sections (cr. upper and lower panels) the basins of attraction occupy essentially different regions although, in general, they are qualitatively similar. This scenario suggests an essential dependence of the global system dynamics of model~(\ref{PN}) on the initial conditions. The shape of basins develope as $\alpha$ approaches $\pi/2$, becoming even "puzzling" as $\alpha$ crosses over $\pi/2$ \cite{brezetsky2021chimera}.

\section{\label{cn}Conclusions}

In conclusion, we have shown that mixed-mode chimera states naturally arise in a small network of $N=3$ coupled pendula with global phase-lagged coupling, and demonstrated the similar but much more developed behaviors of this kind in large network of $N=100$ pendula. Properties of the mixed-mode chimeras specify their analogy to the bump states from neuroscience, both arising as a result of the coupling excitability. With a difference, however, that the mixed-mode chimeras grow in networks with bistable individual dynamics above the Tricomi bifurcation curve. The huge variability of coexisting chimera states indicates, we suggest, an apparent transition to space chaos in model~(\ref{PN}) as $N$ tends to infinity. A detailed verification of this fact can be a subject of future study. It would also be interesting to examine deeper the qualitative connections between the pendula and the neuronal networks, different at the first glance but generating similar patterns. Additionally, the system dynamics in our case is not ruled to a large extend by self-organizing processes (forming rather limited numbers of coherent structures), but by the other ''space chaos'' mechanism in which a combinatorially large variety of the stable states can be created given an appropriate choice of the initial conditions. We believe that the emergence and multiplicity of the mixed-mode chimera states indicates a common, probably universal phenomenon in networks of very different nature, due to the competition between inertia and excitability.

\begin{acknowledgments}
P. Ebrahimzadeh has been supported by the Helmholtz Association Initiative and Networking Fund under project number SO-092 (Advanced Computing
Architectures, ACA) and HITEC program.
\end{acknowledgments}

\appendix*

\section{Asymptotic solution of pendulum network}
\label{AP1}
In order to construct an asymptotic description of model~(\ref{PN}), we introduce a re-scaled time $\tau = \frac{w}{\eps}$ and $\lambda \kappa = \frac{\eps^2}{w^2}$ for some dummy variable $w$. The equation of motion of the network then reads
\bea
\label{a1}
m \phi_i ^{\prime \prime} + \lambda \kappa w \phi_i ^{\prime } + \lambda \kappa r \sin{\phi_i} &=& - \lambda \kappa n_i \tilde{w_i} \\
&+& \lambda \kappa \mu \sum_{j} n_j \sin[\phi_j - \phi_i - \alpha] \nonumber,
\eea
where the prime symbol denotes derivative with respect to re-scaled time ${}^\prime = \frac{\dm}{\dm \tau}$. One can decompose $\phi$ as a monotonic and a perturbebd bounded function with small perturbation $\phi_i \approx \phi_{0,i} + \lambda \dd\phi_{1,i}$. Inserting this condition into Eq.~\ref{a1} and sorting for perturbation parameter $\lambda$ one gets equations of motion for the main and perturbed parts
\bea
\phi^{\prime \prime}_{0,i} &=& 0, \label{a21} \\ 
m \dd \phi^{\prime \prime}_{1,i} + \kappa w \phi^{\prime}_{0,i} + \kappa r \sin{\phi_{0,i}} &=&  -\kappa n_i \tilde{w_i}  \\ \label{a22}
 &+& \kappa \mu \sum_{j} n_j \sin[\phi_{0,j} - \phi_{0,i} - \alpha] \nonumber.
\eea
Equation~\ref{a21} reveals that the unperturbed part $\phi_0$ is indeed a monotonic function in time and gives the average velocity of the pendulum as a constant. The perturbed part $\dd \phi_1$, however, is a bounded function implying Eq.~(A.3) cannot have a non-zero torque. This can be prevented by equating $ w \phi_{0,i}^{\prime} = -n_i \tilde{w_i}$. Rewriting this equation in original time scale, the dummy parameter $w$ drops and one gets the average velocity of the pendulum
\be
\label{a23}
\tilde \omega_i \equiv \dot \phi_{0,i} = -\frac{n_i \tilde{w_i}}{\eps}.
\ee
Inserting this condition into Eq.~(A.3), one gets the equation of motion for the perturbed part in original time scale
\be
\label{}
\lambda \dd \ddot\phi_{1,i} = -r \sin(\tilde \omega_i t) + \mu \sum_{j} n_j \sin[\tilde\omega_j t - \tilde\omega_i t - \alpha],
\ee 
The right hand side is now only a function of time, and one finds the perturbed function $\dd \phi$ by straightforward integration.

\nocite{*}
\bibliography{Refs}

\providecommand{\noopsort}[1]{}\providecommand{\singleletter}[1]{#1}%
\begin{thebibliography}{44}%
\makeatletter
\providecommand \@ifxundefined [1]{%
 \@ifx{#1\undefined}
}%
\providecommand \@ifnum [1]{%
 \ifnum #1\expandafter \@firstoftwo
 \else \expandafter \@secondoftwo
 \fi
}%
\providecommand \@ifx [1]{%
 \ifx #1\expandafter \@firstoftwo
 \else \expandafter \@secondoftwo
 \fi
}%
\providecommand \natexlab [1]{#1}%
\providecommand \enquote  [1]{``#1''}%
\providecommand \bibnamefont  [1]{#1}%
\providecommand \bibfnamefont [1]{#1}%
\providecommand \citenamefont [1]{#1}%
\providecommand \href@noop [0]{\@secondoftwo}%
\providecommand \href [0]{\begingroup \@sanitize@url \@href}%
\providecommand \@href[1]{\@@startlink{#1}\@@href}%
\providecommand \@@href[1]{\endgroup#1\@@endlink}%
\providecommand \@sanitize@url [0]{\catcode `\\12\catcode `\$12\catcode
  `\&12\catcode `\#12\catcode `\^12\catcode `\_12\catcode `\%12\relax}%
\providecommand \@@startlink[1]{}%
\providecommand \@@endlink[0]{}%
\providecommand \url  [0]{\begingroup\@sanitize@url \@url }%
\providecommand \@url [1]{\endgroup\@href {#1}{\urlprefix }}%
\providecommand \urlprefix  [0]{URL }%
\providecommand \Eprint [0]{\href }%
\providecommand \doibase [0]{https://doi.org/}%
\providecommand \selectlanguage [0]{\@gobble}%
\providecommand \bibinfo  [0]{\@secondoftwo}%
\providecommand \bibfield  [0]{\@secondoftwo}%
\providecommand \translation [1]{[#1]}%
\providecommand \BibitemOpen [0]{}%
\providecommand \bibitemStop [0]{}%
\providecommand \bibitemNoStop [0]{.\EOS\space}%
\providecommand \EOS [0]{\spacefactor3000\relax}%
\providecommand \BibitemShut  [1]{\csname bibitem#1\endcsname}%
\let\auto@bib@innerbib\@empty
\bibitem [{\citenamefont {Kuramoto}(2003)}]{kuramoto2003chemical}%
  \BibitemOpen
  \bibfield  {author} {\bibinfo {author} {\bibfnamefont {Y.}~\bibnamefont
  {Kuramoto}},\ }\href@noop {} {\emph {\bibinfo {title} {Chemical oscillations,
  waves, and turbulence}}}\ (\bibinfo  {publisher} {Courier Corporation},\
  \bibinfo {year} {2003})\BibitemShut {NoStop}%
\bibitem [{\citenamefont {Pikovsky}, \citenamefont {Rosenblum},\ and\
  \citenamefont {Kurths}(2001)}]{pikovsky_rosenblum_kurths_2001}%
  \BibitemOpen
  \bibfield  {author} {\bibinfo {author} {\bibfnamefont {A.}~\bibnamefont
  {Pikovsky}}, \bibinfo {author} {\bibfnamefont {M.}~\bibnamefont
  {Rosenblum}},\ and\ \bibinfo {author} {\bibfnamefont {J.}~\bibnamefont
  {Kurths}},\ }\href {https://doi.org/10.1017/CBO9780511755743} {\emph
  {\bibinfo {title} {Synchronization: A Universal Concept in Nonlinear
  Sciences}}},\ Cambridge Nonlinear Science Series\ (\bibinfo  {publisher}
  {Cambridge University Press},\ \bibinfo {year} {2001})\BibitemShut {NoStop}%
\bibitem [{\citenamefont {Strogatz}(2001)}]{strogatz2001exploring}%
  \BibitemOpen
  \bibfield  {author} {\bibinfo {author} {\bibfnamefont {S.~H.}\ \bibnamefont
  {Strogatz}},\ }\bibfield  {title} {\enquote {\bibinfo {title} {Exploring
  complex networks},}\ }\href@noop {} {\bibfield  {journal} {\bibinfo
  {journal} {nature}\ }\textbf {\bibinfo {volume} {410}},\ \bibinfo {pages}
  {268--276} (\bibinfo {year} {2001})}\BibitemShut {NoStop}%
\bibitem [{\citenamefont {Panaggio}\ and\ \citenamefont
  {Abrams}(2015)}]{Panaggio_2015}%
  \BibitemOpen
  \bibfield  {author} {\bibinfo {author} {\bibfnamefont {M.~J.}\ \bibnamefont
  {Panaggio}}\ and\ \bibinfo {author} {\bibfnamefont {D.~M.}\ \bibnamefont
  {Abrams}},\ }\bibfield  {title} {\enquote {\bibinfo {title} {Chimera states:
  coexistence of coherence and incoherence in networks of coupled
  oscillators},}\ }\href {https://doi.org/10.1088/0951-7715/28/3/r67}
  {\bibfield  {journal} {\bibinfo  {journal} {Nonlinearity}\ }\textbf {\bibinfo
  {volume} {28}},\ \bibinfo {pages} {R67--R87} (\bibinfo {year}
  {2015})}\BibitemShut {NoStop}%
\bibitem [{\citenamefont {Sch{\"o}ll}(2016)}]{scholl2016synchronization}%
  \BibitemOpen
  \bibfield  {author} {\bibinfo {author} {\bibfnamefont {E.}~\bibnamefont
  {Sch{\"o}ll}},\ }\bibfield  {title} {\enquote {\bibinfo {title}
  {Synchronization patterns and chimera states in complex networks: Interplay
  of topology and dynamics},}\ }\href@noop {} {\bibfield  {journal} {\bibinfo
  {journal} {The European Physical Journal Special Topics}\ }\textbf {\bibinfo
  {volume} {225}},\ \bibinfo {pages} {891--919} (\bibinfo {year}
  {2016})}\BibitemShut {NoStop}%
\bibitem [{\citenamefont {Omel’chenko}\ and\ \citenamefont
  {Knobloch}(2019)}]{omel2019chimerapedia}%
  \BibitemOpen
  \bibfield  {author} {\bibinfo {author} {\bibfnamefont {E.}~\bibnamefont
  {Omel’chenko}}\ and\ \bibinfo {author} {\bibfnamefont {E.}~\bibnamefont
  {Knobloch}},\ }\bibfield  {title} {\enquote {\bibinfo {title} {Chimerapedia:
  coherence--incoherence patterns in one, two and three dimensions},}\
  }\href@noop {} {\bibfield  {journal} {\bibinfo  {journal} {New Journal of
  Physics}\ }\textbf {\bibinfo {volume} {21}},\ \bibinfo {pages} {093034}
  (\bibinfo {year} {2019})}\BibitemShut {NoStop}%
\bibitem [{\citenamefont {Zakharova}(2020)}]{zakharova2020chimera}%
  \BibitemOpen
  \bibfield  {author} {\bibinfo {author} {\bibfnamefont {A.}~\bibnamefont
  {Zakharova}},\ }\href@noop {} {\emph {\bibinfo {title} {Chimera Patterns in
  Networks}}}\ (\bibinfo  {publisher} {Springer},\ \bibinfo {year}
  {2020})\BibitemShut {NoStop}%
\bibitem [{\citenamefont {Parastesh}\ \emph {et~al.}(2021)\citenamefont
  {Parastesh}, \citenamefont {Jafari}, \citenamefont {Azarnoush}, \citenamefont
  {Shahriari}, \citenamefont {Wang}, \citenamefont {Boccaletti},\ and\
  \citenamefont {Perc}}]{parastesh2021chimeras}%
  \BibitemOpen
  \bibfield  {author} {\bibinfo {author} {\bibfnamefont {F.}~\bibnamefont
  {Parastesh}}, \bibinfo {author} {\bibfnamefont {S.}~\bibnamefont {Jafari}},
  \bibinfo {author} {\bibfnamefont {H.}~\bibnamefont {Azarnoush}}, \bibinfo
  {author} {\bibfnamefont {Z.}~\bibnamefont {Shahriari}}, \bibinfo {author}
  {\bibfnamefont {Z.}~\bibnamefont {Wang}}, \bibinfo {author} {\bibfnamefont
  {S.}~\bibnamefont {Boccaletti}},\ and\ \bibinfo {author} {\bibfnamefont
  {M.}~\bibnamefont {Perc}},\ }\bibfield  {title} {\enquote {\bibinfo {title}
  {Chimeras},}\ }\href@noop {} {\bibfield  {journal} {\bibinfo  {journal}
  {Physics Reports}\ }\textbf {\bibinfo {volume} {898}},\ \bibinfo {pages}
  {1--114} (\bibinfo {year} {2021})}\BibitemShut {NoStop}%
\bibitem [{\citenamefont {Ashwin}\ and\ \citenamefont
  {Burylko}(2015)}]{ashwin2015weak}%
  \BibitemOpen
  \bibfield  {author} {\bibinfo {author} {\bibfnamefont {P.}~\bibnamefont
  {Ashwin}}\ and\ \bibinfo {author} {\bibfnamefont {O.}~\bibnamefont
  {Burylko}},\ }\bibfield  {title} {\enquote {\bibinfo {title} {Weak chimeras
  in minimal networks of coupled phase oscillators},}\ }\href@noop {}
  {\bibfield  {journal} {\bibinfo  {journal} {Chaos: An Interdisciplinary
  Journal of Nonlinear Science}\ }\textbf {\bibinfo {volume} {25}},\ \bibinfo
  {pages} {013106} (\bibinfo {year} {2015})}\BibitemShut {NoStop}%
\bibitem [{\citenamefont {Jaros}\ \emph {et~al.}(2018)\citenamefont {Jaros},
  \citenamefont {Brezetsky}, \citenamefont {Levchenko}, \citenamefont
  {Dudkowski}, \citenamefont {Kapitaniak},\ and\ \citenamefont
  {Maistrenko}}]{jaros2018solitary}%
  \BibitemOpen
  \bibfield  {author} {\bibinfo {author} {\bibfnamefont {P.}~\bibnamefont
  {Jaros}}, \bibinfo {author} {\bibfnamefont {S.}~\bibnamefont {Brezetsky}},
  \bibinfo {author} {\bibfnamefont {R.}~\bibnamefont {Levchenko}}, \bibinfo
  {author} {\bibfnamefont {D.}~\bibnamefont {Dudkowski}}, \bibinfo {author}
  {\bibfnamefont {T.}~\bibnamefont {Kapitaniak}},\ and\ \bibinfo {author}
  {\bibfnamefont {Y.}~\bibnamefont {Maistrenko}},\ }\bibfield  {title}
  {\enquote {\bibinfo {title} {Solitary states for coupled oscillators with
  inertia},}\ }\href@noop {} {\bibfield  {journal} {\bibinfo  {journal} {Chaos:
  An Interdisciplinary Journal of Nonlinear Science}\ }\textbf {\bibinfo
  {volume} {28}},\ \bibinfo {pages} {011103} (\bibinfo {year}
  {2018})}\BibitemShut {NoStop}%
\bibitem [{\citenamefont {Berner}\ \emph {et~al.}(2020)\citenamefont {Berner},
  \citenamefont {Polanska}, \citenamefont {Sch{\"o}ll},\ and\ \citenamefont
  {Yanchuk}}]{berner2020solitary}%
  \BibitemOpen
  \bibfield  {author} {\bibinfo {author} {\bibfnamefont {R.}~\bibnamefont
  {Berner}}, \bibinfo {author} {\bibfnamefont {A.}~\bibnamefont {Polanska}},
  \bibinfo {author} {\bibfnamefont {E.}~\bibnamefont {Sch{\"o}ll}},\ and\
  \bibinfo {author} {\bibfnamefont {S.}~\bibnamefont {Yanchuk}},\ }\bibfield
  {title} {\enquote {\bibinfo {title} {Solitary states in adaptive nonlocal
  oscillator networks},}\ }\href@noop {} {\bibfield  {journal} {\bibinfo
  {journal} {The European Physical Journal Special Topics}\ }\textbf {\bibinfo
  {volume} {229}},\ \bibinfo {pages} {2183--2203} (\bibinfo {year}
  {2020})}\BibitemShut {NoStop}%
\bibitem [{\citenamefont {Jaros}\ \emph {et~al.}(2021)\citenamefont {Jaros},
  \citenamefont {Levchenko}, \citenamefont {Kapitaniak},\ and\ \citenamefont
  {Maistrenko}}]{jaros2021chimera}%
  \BibitemOpen
  \bibfield  {author} {\bibinfo {author} {\bibfnamefont {P.}~\bibnamefont
  {Jaros}}, \bibinfo {author} {\bibfnamefont {R.}~\bibnamefont {Levchenko}},
  \bibinfo {author} {\bibfnamefont {T.}~\bibnamefont {Kapitaniak}},\ and\
  \bibinfo {author} {\bibfnamefont {Y.}~\bibnamefont {Maistrenko}},\ }\bibfield
   {title} {\enquote {\bibinfo {title} {Chimera states for directed
  networks},}\ }\href@noop {} {\bibfield  {journal} {\bibinfo  {journal}
  {Chaos: An Interdisciplinary Journal of Nonlinear Science}\ }\textbf
  {\bibinfo {volume} {31}},\ \bibinfo {pages} {103111} (\bibinfo {year}
  {2021})}\BibitemShut {NoStop}%
\bibitem [{\citenamefont {Sch{\"u}len}\ \emph {et~al.}(2021)\citenamefont
  {Sch{\"u}len}, \citenamefont {Janzen}, \citenamefont {Medeiros},\ and\
  \citenamefont {Zakharova}}]{schulen2021solitary}%
  \BibitemOpen
  \bibfield  {author} {\bibinfo {author} {\bibfnamefont {L.}~\bibnamefont
  {Sch{\"u}len}}, \bibinfo {author} {\bibfnamefont {D.~A.}\ \bibnamefont
  {Janzen}}, \bibinfo {author} {\bibfnamefont {E.~S.}\ \bibnamefont
  {Medeiros}},\ and\ \bibinfo {author} {\bibfnamefont {A.}~\bibnamefont
  {Zakharova}},\ }\bibfield  {title} {\enquote {\bibinfo {title} {Solitary
  states in multiplex neural networks: Onset and vulnerability},}\ }\href@noop
  {} {\bibfield  {journal} {\bibinfo  {journal} {Chaos, Solitons \& Fractals}\
  }\textbf {\bibinfo {volume} {145}},\ \bibinfo {pages} {110670} (\bibinfo
  {year} {2021})}\BibitemShut {NoStop}%
\bibitem [{\citenamefont {Hellmann}\ \emph {et~al.}(2020)\citenamefont
  {Hellmann}, \citenamefont {Schultz}, \citenamefont {Jaros}, \citenamefont
  {Levchenko}, \citenamefont {Kapitaniak}, \citenamefont {Kurths},\ and\
  \citenamefont {Maistrenko}}]{hellmann2020network}%
  \BibitemOpen
  \bibfield  {author} {\bibinfo {author} {\bibfnamefont {F.}~\bibnamefont
  {Hellmann}}, \bibinfo {author} {\bibfnamefont {P.}~\bibnamefont {Schultz}},
  \bibinfo {author} {\bibfnamefont {P.}~\bibnamefont {Jaros}}, \bibinfo
  {author} {\bibfnamefont {R.}~\bibnamefont {Levchenko}}, \bibinfo {author}
  {\bibfnamefont {T.}~\bibnamefont {Kapitaniak}}, \bibinfo {author}
  {\bibfnamefont {J.}~\bibnamefont {Kurths}},\ and\ \bibinfo {author}
  {\bibfnamefont {Y.}~\bibnamefont {Maistrenko}},\ }\bibfield  {title}
  {\enquote {\bibinfo {title} {Network-induced multistability through lossy
  coupling and exotic solitary states},}\ }\href@noop {} {\bibfield  {journal}
  {\bibinfo  {journal} {Nature communications}\ }\textbf {\bibinfo {volume}
  {11}},\ \bibinfo {pages} {1--9} (\bibinfo {year} {2020})}\BibitemShut
  {NoStop}%
\bibitem [{\citenamefont {Franovi{\'c}}\ \emph {et~al.}(2022)\citenamefont
  {Franovi{\'c}}, \citenamefont {Eydam}, \citenamefont {Semenova},\ and\
  \citenamefont {Zakharova}}]{franovic2022unbalanced}%
  \BibitemOpen
  \bibfield  {author} {\bibinfo {author} {\bibfnamefont {I.}~\bibnamefont
  {Franovi{\'c}}}, \bibinfo {author} {\bibfnamefont {S.}~\bibnamefont {Eydam}},
  \bibinfo {author} {\bibfnamefont {N.}~\bibnamefont {Semenova}},\ and\
  \bibinfo {author} {\bibfnamefont {A.}~\bibnamefont {Zakharova}},\ }\bibfield
  {title} {\enquote {\bibinfo {title} {Unbalanced clustering and solitary
  states in coupled excitable systems},}\ }\href@noop {} {\bibfield  {journal}
  {\bibinfo  {journal} {Chaos: An Interdisciplinary Journal of Nonlinear
  Science}\ }\textbf {\bibinfo {volume} {32}},\ \bibinfo {pages} {011104}
  (\bibinfo {year} {2022})}\BibitemShut {NoStop}%
\bibitem [{\citenamefont {Maistrenko}, \citenamefont {Penkovsky},\ and\
  \citenamefont {Rosenblum}(2014)}]{maistrenko2014solitary}%
  \BibitemOpen
  \bibfield  {author} {\bibinfo {author} {\bibfnamefont {Y.}~\bibnamefont
  {Maistrenko}}, \bibinfo {author} {\bibfnamefont {B.}~\bibnamefont
  {Penkovsky}},\ and\ \bibinfo {author} {\bibfnamefont {M.}~\bibnamefont
  {Rosenblum}},\ }\bibfield  {title} {\enquote {\bibinfo {title} {Solitary
  state at the edge of synchrony in ensembles with attractive and repulsive
  interactions},}\ }\href@noop {} {\bibfield  {journal} {\bibinfo  {journal}
  {Physical Review E}\ }\textbf {\bibinfo {volume} {89}},\ \bibinfo {pages}
  {060901} (\bibinfo {year} {2014})}\BibitemShut {NoStop}%
\bibitem [{\citenamefont {Maistrenko}, \citenamefont {Sudakov},\ and\
  \citenamefont {Maistrenko}(2020)}]{maistrenko2020spiral}%
  \BibitemOpen
  \bibfield  {author} {\bibinfo {author} {\bibfnamefont {V.}~\bibnamefont
  {Maistrenko}}, \bibinfo {author} {\bibfnamefont {O.}~\bibnamefont
  {Sudakov}},\ and\ \bibinfo {author} {\bibfnamefont {Y.}~\bibnamefont
  {Maistrenko}},\ }\bibfield  {title} {\enquote {\bibinfo {title} {Spiral wave
  chimeras for coupled oscillators with inertia},}\ }\href@noop {} {\bibfield
  {journal} {\bibinfo  {journal} {The European Physical Journal Special
  Topics}\ }\textbf {\bibinfo {volume} {229}},\ \bibinfo {pages} {2327--2340}
  (\bibinfo {year} {2020})}\BibitemShut {NoStop}%
\bibitem [{\citenamefont {Maistrenko}, \citenamefont {Sudakov},\ and\
  \citenamefont {Osiv}(2020)}]{maistrenko2020chimeras}%
  \BibitemOpen
  \bibfield  {author} {\bibinfo {author} {\bibfnamefont {V.}~\bibnamefont
  {Maistrenko}}, \bibinfo {author} {\bibfnamefont {O.}~\bibnamefont
  {Sudakov}},\ and\ \bibinfo {author} {\bibfnamefont {O.}~\bibnamefont
  {Osiv}},\ }\bibfield  {title} {\enquote {\bibinfo {title} {Chimeras and
  solitary states in 3d oscillator networks with inertia},}\ }\href@noop {}
  {\bibfield  {journal} {\bibinfo  {journal} {Chaos: An Interdisciplinary
  Journal of Nonlinear Science}\ }\textbf {\bibinfo {volume} {30}},\ \bibinfo
  {pages} {063113} (\bibinfo {year} {2020})}\BibitemShut {NoStop}%
\bibitem [{\citenamefont {Jaros}, \citenamefont {Maistrenko},\ and\
  \citenamefont {Kapitaniak}(2015)}]{jaros2015chimera}%
  \BibitemOpen
  \bibfield  {author} {\bibinfo {author} {\bibfnamefont {P.}~\bibnamefont
  {Jaros}}, \bibinfo {author} {\bibfnamefont {Y.}~\bibnamefont {Maistrenko}},\
  and\ \bibinfo {author} {\bibfnamefont {T.}~\bibnamefont {Kapitaniak}},\
  }\bibfield  {title} {\enquote {\bibinfo {title} {Chimera states on the route
  from coherence to rotating waves},}\ }\href@noop {} {\bibfield  {journal}
  {\bibinfo  {journal} {Physical Review E}\ }\textbf {\bibinfo {volume} {91}},\
  \bibinfo {pages} {022907} (\bibinfo {year} {2015})}\BibitemShut {NoStop}%
\bibitem [{\citenamefont {Ermentrout}(1991)}]{ermentrout1991adaptive}%
  \BibitemOpen
  \bibfield  {author} {\bibinfo {author} {\bibfnamefont {B.}~\bibnamefont
  {Ermentrout}},\ }\bibfield  {title} {\enquote {\bibinfo {title} {An adaptive
  model for synchrony in the firefly pteroptyx malaccae},}\ }\href@noop {}
  {\bibfield  {journal} {\bibinfo  {journal} {Journal of Mathematical Biology}\
  }\textbf {\bibinfo {volume} {29}},\ \bibinfo {pages} {571--585} (\bibinfo
  {year} {1991})}\BibitemShut {NoStop}%
\bibitem [{\citenamefont {Olmi}\ \emph {et~al.}(2015)\citenamefont {Olmi},
  \citenamefont {Martens}, \citenamefont {Thutupalli},\ and\ \citenamefont
  {Torcini}}]{olmi2015intermittent}%
  \BibitemOpen
  \bibfield  {author} {\bibinfo {author} {\bibfnamefont {S.}~\bibnamefont
  {Olmi}}, \bibinfo {author} {\bibfnamefont {E.~A.}\ \bibnamefont {Martens}},
  \bibinfo {author} {\bibfnamefont {S.}~\bibnamefont {Thutupalli}},\ and\
  \bibinfo {author} {\bibfnamefont {A.}~\bibnamefont {Torcini}},\ }\bibfield
  {title} {\enquote {\bibinfo {title} {Intermittent chaotic chimeras for
  coupled rotators},}\ }\href@noop {} {\bibfield  {journal} {\bibinfo
  {journal} {Physical Review E}\ }\textbf {\bibinfo {volume} {92}},\ \bibinfo
  {pages} {030901} (\bibinfo {year} {2015})}\BibitemShut {NoStop}%
\bibitem [{\citenamefont {Olmi}(2015)}]{olmi2015chimera}%
  \BibitemOpen
  \bibfield  {author} {\bibinfo {author} {\bibfnamefont {S.}~\bibnamefont
  {Olmi}},\ }\bibfield  {title} {\enquote {\bibinfo {title} {Chimera states in
  coupled kuramoto oscillators with inertia},}\ }\href@noop {} {\bibfield
  {journal} {\bibinfo  {journal} {Chaos: An Interdisciplinary Journal of
  Nonlinear Science}\ }\textbf {\bibinfo {volume} {25}},\ \bibinfo {pages}
  {123125} (\bibinfo {year} {2015})}\BibitemShut {NoStop}%
\bibitem [{\citenamefont {Belykh}, \citenamefont {Brister},\ and\ \citenamefont
  {Belykh}(2016)}]{belykh2016bistability}%
  \BibitemOpen
  \bibfield  {author} {\bibinfo {author} {\bibfnamefont {I.~V.}\ \bibnamefont
  {Belykh}}, \bibinfo {author} {\bibfnamefont {B.~N.}\ \bibnamefont
  {Brister}},\ and\ \bibinfo {author} {\bibfnamefont {V.~N.}\ \bibnamefont
  {Belykh}},\ }\bibfield  {title} {\enquote {\bibinfo {title} {Bistability of
  patterns of synchrony in kuramoto oscillators with inertia},}\ }\href@noop {}
  {\bibfield  {journal} {\bibinfo  {journal} {Chaos: An Interdisciplinary
  Journal of Nonlinear Science}\ }\textbf {\bibinfo {volume} {26}},\ \bibinfo
  {pages} {094822} (\bibinfo {year} {2016})}\BibitemShut {NoStop}%
\bibitem [{\citenamefont {Belykh}, \citenamefont {Jeter},\ and\ \citenamefont
  {Belykh}(2017)}]{belykh2017foot}%
  \BibitemOpen
  \bibfield  {author} {\bibinfo {author} {\bibfnamefont {I.}~\bibnamefont
  {Belykh}}, \bibinfo {author} {\bibfnamefont {R.}~\bibnamefont {Jeter}},\ and\
  \bibinfo {author} {\bibfnamefont {V.}~\bibnamefont {Belykh}},\ }\bibfield
  {title} {\enquote {\bibinfo {title} {Foot force models of crowd dynamics on a
  wobbly bridge},}\ }\href@noop {} {\bibfield  {journal} {\bibinfo  {journal}
  {Science advances}\ }\textbf {\bibinfo {volume} {3}},\ \bibinfo {pages}
  {e1701512} (\bibinfo {year} {2017})}\BibitemShut {NoStop}%
\bibitem [{\citenamefont {Brister}, \citenamefont {Belykh},\ and\ \citenamefont
  {Belykh}(2020)}]{brister2020three}%
  \BibitemOpen
  \bibfield  {author} {\bibinfo {author} {\bibfnamefont {B.~N.}\ \bibnamefont
  {Brister}}, \bibinfo {author} {\bibfnamefont {V.~N.}\ \bibnamefont
  {Belykh}},\ and\ \bibinfo {author} {\bibfnamefont {I.~V.}\ \bibnamefont
  {Belykh}},\ }\bibfield  {title} {\enquote {\bibinfo {title} {When three is a
  crowd: Chaos from clusters of kuramoto oscillators with inertia},}\
  }\href@noop {} {\bibfield  {journal} {\bibinfo  {journal} {Physical Review
  E}\ }\textbf {\bibinfo {volume} {101}},\ \bibinfo {pages} {062206} (\bibinfo
  {year} {2020})}\BibitemShut {NoStop}%
\bibitem [{\citenamefont {Kruk}, \citenamefont {Maistrenko},\ and\
  \citenamefont {Koeppl}(2020)}]{kruk2020solitary}%
  \BibitemOpen
  \bibfield  {author} {\bibinfo {author} {\bibfnamefont {N.}~\bibnamefont
  {Kruk}}, \bibinfo {author} {\bibfnamefont {Y.}~\bibnamefont {Maistrenko}},\
  and\ \bibinfo {author} {\bibfnamefont {H.}~\bibnamefont {Koeppl}},\
  }\bibfield  {title} {\enquote {\bibinfo {title} {Solitary states in the
  mean-field limit},}\ }\href@noop {} {\bibfield  {journal} {\bibinfo
  {journal} {Chaos: An Interdisciplinary Journal of Nonlinear Science}\
  }\textbf {\bibinfo {volume} {30}},\ \bibinfo {pages} {111104} (\bibinfo
  {year} {2020})}\BibitemShut {NoStop}%
\bibitem [{\citenamefont {Kuramoto}\ and\ \citenamefont
  {Battogtokh}(2002)}]{kuramoto2002coexistence}%
  \BibitemOpen
  \bibfield  {author} {\bibinfo {author} {\bibfnamefont {Y.}~\bibnamefont
  {Kuramoto}}\ and\ \bibinfo {author} {\bibfnamefont {D.}~\bibnamefont
  {Battogtokh}},\ }\bibfield  {title} {\enquote {\bibinfo {title} {Coexistence
  of coherence and incoherence in nonlocally coupled phase oscillators.}}\
  }\href@noop {} {\bibfield  {journal} {\bibinfo  {journal} {NONLINEAR
  PHENOMENA IN COMPLEX SYSTEMS}\ }\textbf {\bibinfo {volume} {5}},\ \bibinfo
  {pages} {380--385} (\bibinfo {year} {2002})}\BibitemShut {NoStop}%
\bibitem [{\citenamefont {Abrams}\ and\ \citenamefont
  {Strogatz}(2004)}]{PhysRevLett.93.174102}%
  \BibitemOpen
  \bibfield  {author} {\bibinfo {author} {\bibfnamefont {D.~M.}\ \bibnamefont
  {Abrams}}\ and\ \bibinfo {author} {\bibfnamefont {S.~H.}\ \bibnamefont
  {Strogatz}},\ }\bibfield  {title} {\enquote {\bibinfo {title} {Chimera states
  for coupled oscillators},}\ }\href
  {https://doi.org/10.1103/PhysRevLett.93.174102} {\bibfield  {journal}
  {\bibinfo  {journal} {Phys. Rev. Lett.}\ }\textbf {\bibinfo {volume} {93}},\
  \bibinfo {pages} {174102} (\bibinfo {year} {2004})}\BibitemShut {NoStop}%
\bibitem [{\citenamefont {Omelchenko}\ \emph {et~al.}(2011)\citenamefont
  {Omelchenko}, \citenamefont {Maistrenko}, \citenamefont {H{\"o}vel},\ and\
  \citenamefont {Sch{\"o}ll}}]{omelchenko2011loss}%
  \BibitemOpen
  \bibfield  {author} {\bibinfo {author} {\bibfnamefont {I.}~\bibnamefont
  {Omelchenko}}, \bibinfo {author} {\bibfnamefont {Y.}~\bibnamefont
  {Maistrenko}}, \bibinfo {author} {\bibfnamefont {P.}~\bibnamefont
  {H{\"o}vel}},\ and\ \bibinfo {author} {\bibfnamefont {E.}~\bibnamefont
  {Sch{\"o}ll}},\ }\bibfield  {title} {\enquote {\bibinfo {title} {Loss of
  coherence in dynamical networks: spatial chaos and chimera states},}\
  }\href@noop {} {\bibfield  {journal} {\bibinfo  {journal} {Physical review
  letters}\ }\textbf {\bibinfo {volume} {106}},\ \bibinfo {pages} {234102}
  (\bibinfo {year} {2011})}\BibitemShut {NoStop}%
\bibitem [{\citenamefont {Kapitaniak}\ \emph {et~al.}(2014)\citenamefont
  {Kapitaniak}, \citenamefont {Kuzma}, \citenamefont {Wojewoda}, \citenamefont
  {Czolczynski},\ and\ \citenamefont {Maistrenko}}]{kapitaniak2014imperfect}%
  \BibitemOpen
  \bibfield  {author} {\bibinfo {author} {\bibfnamefont {T.}~\bibnamefont
  {Kapitaniak}}, \bibinfo {author} {\bibfnamefont {P.}~\bibnamefont {Kuzma}},
  \bibinfo {author} {\bibfnamefont {J.}~\bibnamefont {Wojewoda}}, \bibinfo
  {author} {\bibfnamefont {K.}~\bibnamefont {Czolczynski}},\ and\ \bibinfo
  {author} {\bibfnamefont {Y.}~\bibnamefont {Maistrenko}},\ }\bibfield  {title}
  {\enquote {\bibinfo {title} {Imperfect chimera states for coupled pendula},}\
  }\href@noop {} {\bibfield  {journal} {\bibinfo  {journal} {Scientific
  reports}\ }\textbf {\bibinfo {volume} {4}},\ \bibinfo {pages} {1--4}
  (\bibinfo {year} {2014})}\BibitemShut {NoStop}%
\bibitem [{\citenamefont {Wojewoda}\ \emph {et~al.}(2016)\citenamefont
  {Wojewoda}, \citenamefont {Czolczynski}, \citenamefont {Maistrenko},\ and\
  \citenamefont {Kapitaniak}}]{wojewoda2016smallest}%
  \BibitemOpen
  \bibfield  {author} {\bibinfo {author} {\bibfnamefont {J.}~\bibnamefont
  {Wojewoda}}, \bibinfo {author} {\bibfnamefont {K.}~\bibnamefont
  {Czolczynski}}, \bibinfo {author} {\bibfnamefont {Y.}~\bibnamefont
  {Maistrenko}},\ and\ \bibinfo {author} {\bibfnamefont {T.}~\bibnamefont
  {Kapitaniak}},\ }\bibfield  {title} {\enquote {\bibinfo {title} {The smallest
  chimera state for coupled pendula},}\ }\href@noop {} {\bibfield  {journal}
  {\bibinfo  {journal} {Scientific reports}\ }\textbf {\bibinfo {volume} {6}},\
  \bibinfo {pages} {1--5} (\bibinfo {year} {2016})}\BibitemShut {NoStop}%
\bibitem [{\citenamefont {Ebrahimzadeh}\ \emph {et~al.}(2020)\citenamefont
  {Ebrahimzadeh}, \citenamefont {Schiek}, \citenamefont {Jaros}, \citenamefont
  {Kapitaniak}, \citenamefont {van Waasen},\ and\ \citenamefont
  {Maistrenko}}]{ebrahimzadeh2020minimal}%
  \BibitemOpen
  \bibfield  {author} {\bibinfo {author} {\bibfnamefont {P.}~\bibnamefont
  {Ebrahimzadeh}}, \bibinfo {author} {\bibfnamefont {M.}~\bibnamefont
  {Schiek}}, \bibinfo {author} {\bibfnamefont {P.}~\bibnamefont {Jaros}},
  \bibinfo {author} {\bibfnamefont {T.}~\bibnamefont {Kapitaniak}}, \bibinfo
  {author} {\bibfnamefont {S.}~\bibnamefont {van Waasen}},\ and\ \bibinfo
  {author} {\bibfnamefont {Y.}~\bibnamefont {Maistrenko}},\ }\bibfield  {title}
  {\enquote {\bibinfo {title} {Minimal chimera states in phase-lag coupled
  mechanical oscillators},}\ }\href@noop {} {\bibfield  {journal} {\bibinfo
  {journal} {The European Physical Journal Special Topics}\ }\textbf {\bibinfo
  {volume} {229}},\ \bibinfo {pages} {2205--2214} (\bibinfo {year}
  {2020})}\BibitemShut {NoStop}%
\bibitem [{\citenamefont {Brezetsky}\ \emph {et~al.}(2021)\citenamefont
  {Brezetsky}, \citenamefont {Jaros}, \citenamefont {Levchenko}, \citenamefont
  {Kapitaniak},\ and\ \citenamefont {Maistrenko}}]{brezetsky2021chimera}%
  \BibitemOpen
  \bibfield  {author} {\bibinfo {author} {\bibfnamefont {S.}~\bibnamefont
  {Brezetsky}}, \bibinfo {author} {\bibfnamefont {P.}~\bibnamefont {Jaros}},
  \bibinfo {author} {\bibfnamefont {R.}~\bibnamefont {Levchenko}}, \bibinfo
  {author} {\bibfnamefont {T.}~\bibnamefont {Kapitaniak}},\ and\ \bibinfo
  {author} {\bibfnamefont {Y.}~\bibnamefont {Maistrenko}},\ }\bibfield  {title}
  {\enquote {\bibinfo {title} {Chimera complexity},}\ }\href@noop {} {\bibfield
   {journal} {\bibinfo  {journal} {Physical Review E}\ }\textbf {\bibinfo
  {volume} {103}},\ \bibinfo {pages} {L050204} (\bibinfo {year}
  {2021})}\BibitemShut {NoStop}%
\bibitem [{\citenamefont {Ermentrout}(1998)}]{Ermentrout_1998}%
  \BibitemOpen
  \bibfield  {author} {\bibinfo {author} {\bibfnamefont {B.}~\bibnamefont
  {Ermentrout}},\ }\bibfield  {title} {\enquote {\bibinfo {title} {Neural
  networks as spatio-temporal pattern-forming systems},}\ }\href
  {https://doi.org/10.1088/0034-4885/61/4/002} {\bibfield  {journal} {\bibinfo
  {journal} {Reports on Progress in Physics}\ }\textbf {\bibinfo {volume}
  {61}},\ \bibinfo {pages} {353--430} (\bibinfo {year} {1998})}\BibitemShut
  {NoStop}%
\bibitem [{\citenamefont {Laing}\ and\ \citenamefont
  {Chow}(2001)}]{laing2001stationary}%
  \BibitemOpen
  \bibfield  {author} {\bibinfo {author} {\bibfnamefont {C.~R.}\ \bibnamefont
  {Laing}}\ and\ \bibinfo {author} {\bibfnamefont {C.~C.}\ \bibnamefont
  {Chow}},\ }\bibfield  {title} {\enquote {\bibinfo {title} {Stationary bumps
  in networks of spiking neurons},}\ }\href@noop {} {\bibfield  {journal}
  {\bibinfo  {journal} {Neural computation}\ }\textbf {\bibinfo {volume}
  {13}},\ \bibinfo {pages} {1473--1494} (\bibinfo {year} {2001})}\BibitemShut
  {NoStop}%
\bibitem [{\citenamefont {Laing}\ \emph {et~al.}(2002)\citenamefont {Laing},
  \citenamefont {Troy}, \citenamefont {Gutkin},\ and\ \citenamefont
  {Ermentrout}}]{laing2002multiple}%
  \BibitemOpen
  \bibfield  {author} {\bibinfo {author} {\bibfnamefont {C.~R.}\ \bibnamefont
  {Laing}}, \bibinfo {author} {\bibfnamefont {W.~C.}\ \bibnamefont {Troy}},
  \bibinfo {author} {\bibfnamefont {B.}~\bibnamefont {Gutkin}},\ and\ \bibinfo
  {author} {\bibfnamefont {G.~B.}\ \bibnamefont {Ermentrout}},\ }\bibfield
  {title} {\enquote {\bibinfo {title} {Multiple bumps in a neuronal model of
  working memory},}\ }\href@noop {} {\bibfield  {journal} {\bibinfo  {journal}
  {SIAM Journal on Applied Mathematics}\ }\textbf {\bibinfo {volume} {63}},\
  \bibinfo {pages} {62--97} (\bibinfo {year} {2002})}\BibitemShut {NoStop}%
\bibitem [{\citenamefont {Owen}, \citenamefont {Laing},\ and\ \citenamefont
  {Coombes}(2007)}]{owen2007bumps}%
  \BibitemOpen
  \bibfield  {author} {\bibinfo {author} {\bibfnamefont {M.~R.}\ \bibnamefont
  {Owen}}, \bibinfo {author} {\bibfnamefont {C.~R.}\ \bibnamefont {Laing}},\
  and\ \bibinfo {author} {\bibfnamefont {S.}~\bibnamefont {Coombes}},\
  }\bibfield  {title} {\enquote {\bibinfo {title} {Bumps and rings in a
  two-dimensional neural field: splitting and rotational instabilities},}\
  }\href@noop {} {\bibfield  {journal} {\bibinfo  {journal} {New Journal of
  Physics}\ }\textbf {\bibinfo {volume} {9}},\ \bibinfo {pages} {378} (\bibinfo
  {year} {2007})}\BibitemShut {NoStop}%
\bibitem [{\citenamefont {Laing}(2021)}]{laing2021interpolating}%
  \BibitemOpen
  \bibfield  {author} {\bibinfo {author} {\bibfnamefont {C.~R.}\ \bibnamefont
  {Laing}},\ }\bibfield  {title} {\enquote {\bibinfo {title} {Interpolating
  between bumps and chimeras},}\ }\href@noop {} {\bibfield  {journal} {\bibinfo
   {journal} {Chaos: An Interdisciplinary Journal of Nonlinear Science}\
  }\textbf {\bibinfo {volume} {31}},\ \bibinfo {pages} {113116} (\bibinfo
  {year} {2021})}\BibitemShut {NoStop}%
\bibitem [{\citenamefont {Renart}, \citenamefont {Song},\ and\ \citenamefont
  {Wang}(2003)}]{RENART2003473}%
  \BibitemOpen
  \bibfield  {author} {\bibinfo {author} {\bibfnamefont {A.}~\bibnamefont
  {Renart}}, \bibinfo {author} {\bibfnamefont {P.}~\bibnamefont {Song}},\ and\
  \bibinfo {author} {\bibfnamefont {X.-J.}\ \bibnamefont {Wang}},\ }\bibfield
  {title} {\enquote {\bibinfo {title} {Robust spatial working memory through
  homeostatic synaptic scaling in heterogeneous cortical networks},}\ }\href
  {https://doi.org/https://doi.org/10.1016/S0896-6273(03)00255-1} {\bibfield
  {journal} {\bibinfo  {journal} {Neuron}\ }\textbf {\bibinfo {volume} {38}},\
  \bibinfo {pages} {473--485} (\bibinfo {year} {2003})}\BibitemShut {NoStop}%
\bibitem [{\citenamefont {Coombes}(2005)}]{coombes2005waves}%
  \BibitemOpen
  \bibfield  {author} {\bibinfo {author} {\bibfnamefont {S.}~\bibnamefont
  {Coombes}},\ }\bibfield  {title} {\enquote {\bibinfo {title} {Waves, bumps,
  and patterns in neural field theories},}\ }\href@noop {} {\bibfield
  {journal} {\bibinfo  {journal} {Biological cybernetics}\ }\textbf {\bibinfo
  {volume} {93}},\ \bibinfo {pages} {91--108} (\bibinfo {year}
  {2005})}\BibitemShut {NoStop}%
\bibitem [{\citenamefont {Miller}, \citenamefont {Lundqvist},\ and\
  \citenamefont {Bastos}(2018)}]{MILLER2018463}%
  \BibitemOpen
  \bibfield  {author} {\bibinfo {author} {\bibfnamefont {E.~K.}\ \bibnamefont
  {Miller}}, \bibinfo {author} {\bibfnamefont {M.}~\bibnamefont {Lundqvist}},\
  and\ \bibinfo {author} {\bibfnamefont {A.~M.}\ \bibnamefont {Bastos}},\
  }\bibfield  {title} {\enquote {\bibinfo {title} {Working memory 2.0},}\
  }\href {https://doi.org/https://doi.org/10.1016/j.neuron.2018.09.023}
  {\bibfield  {journal} {\bibinfo  {journal} {Neuron}\ }\textbf {\bibinfo
  {volume} {100}},\ \bibinfo {pages} {463--475} (\bibinfo {year}
  {2018})}\BibitemShut {NoStop}%
\bibitem [{\citenamefont {Tricomi}(1933)}]{tricomi1933integrazione}%
  \BibitemOpen
  \bibfield  {author} {\bibinfo {author} {\bibfnamefont {F.}~\bibnamefont
  {Tricomi}},\ }\bibfield  {title} {\enquote {\bibinfo {title} {Integrazione di
  un'equazione differenziale presentatasi in elettrotecnica},}\ }\href@noop {}
  {\bibfield  {journal} {\bibinfo  {journal} {Annali della Scuola Normale
  Superiore di Pisa-Classe di Scienze}\ }\textbf {\bibinfo {volume} {2}},\
  \bibinfo {pages} {1--20} (\bibinfo {year} {1933})}\BibitemShut {NoStop}%
\bibitem [{\citenamefont {Wolfrum}\ and\ \citenamefont
  {Omel’chenko}(2011)}]{wolfrum2011chimera}%
  \BibitemOpen
  \bibfield  {author} {\bibinfo {author} {\bibfnamefont {M.}~\bibnamefont
  {Wolfrum}}\ and\ \bibinfo {author} {\bibfnamefont {E.}~\bibnamefont
  {Omel’chenko}},\ }\bibfield  {title} {\enquote {\bibinfo {title} {Chimera
  states are chaotic transients},}\ }\href@noop {} {\bibfield  {journal}
  {\bibinfo  {journal} {Physical Review E}\ }\textbf {\bibinfo {volume} {84}},\
  \bibinfo {pages} {015201} (\bibinfo {year} {2011})}\BibitemShut {NoStop}%
\bibitem [{\citenamefont {Maistrenko}\ \emph {et~al.}(2017)\citenamefont
  {Maistrenko}, \citenamefont {Brezetsky}, \citenamefont {Jaros}, \citenamefont
  {Levchenko},\ and\ \citenamefont {Kapitaniak}}]{maistrenko2017smallest}%
  \BibitemOpen
  \bibfield  {author} {\bibinfo {author} {\bibfnamefont {Y.}~\bibnamefont
  {Maistrenko}}, \bibinfo {author} {\bibfnamefont {S.}~\bibnamefont
  {Brezetsky}}, \bibinfo {author} {\bibfnamefont {P.}~\bibnamefont {Jaros}},
  \bibinfo {author} {\bibfnamefont {R.}~\bibnamefont {Levchenko}},\ and\
  \bibinfo {author} {\bibfnamefont {T.}~\bibnamefont {Kapitaniak}},\ }\bibfield
   {title} {\enquote {\bibinfo {title} {Smallest chimera states},}\ }\href@noop
  {} {\bibfield  {journal} {\bibinfo  {journal} {Physical Review E}\ }\textbf
  {\bibinfo {volume} {95}},\ \bibinfo {pages} {010203} (\bibinfo {year}
  {2017})}\BibitemShut {NoStop}%
\end{thebibliography}%

\end{document}